\documentclass[12pt]{article}

\usepackage{libertine}
\usepackage{libertinust1math}
\usepackage[T1]{fontenc}

\usepackage{amsmath, amsthm, amsfonts, amssymb}
\usepackage[round]{natbib}
\usepackage[margin=1.5in]{geometry}
\usepackage{setspace}
\usepackage{graphicx, tikzsymbols, parskip, float, array, accents, fancyhdr, dsfont, comment, csquotes}
\usepackage{sectsty, titlesec, subcaption}
\usepackage{bbm, enumitem, bibentry}
\usepackage[utf8]{inputenc}
\definecolor{nightblue}{cmyk}{0, 0.7808, 0.4429, 0.1412}
\usepackage[pagebackref,colorlinks=true,citecolor=nightblue]{hyperref}
\usepackage{indentfirst}
\usepackage{ragged2e}
\sectionfont{\fontsize{13}{15}\selectfont}
\subsectionfont{\fontsize{12}{15}\selectfont}

\theoremstyle{plain}

\usepackage[toc,page]{appendix}
\usepackage{setspace}

\title{\vspace{-2cm} \Large{\bfseries The Moral Burden of Ambiguity Aversion}}

\author{Brian Jabarian\footnote{Paris School of Economics, Panth\'eon-Sorbonne University and Department of Economics, Princeton University. email: \href{jabarian@princeton.edu}{jabarian@princeton.edu}. I thank Franz Dietrich, Marc Fleurbaey, Thomas Rowe, Alex Voorhoeve and Hayden Wilkinson for their written comments. I also thank K\'evin Bertrand, D\'eborah Chekroun, Anders Priergaard, Jimi Vaubien and the audience from the Social Ethics \& Economics Discussion Seminar at Princeton University for useful discussions. I gratefully thank the editors of this PEA Soup P\&PA Discussion, David Faraci and Peter M. Jaworski for their invitation to contribute to this edition. The original version of this paper can be found: \href{http://peasoup.us/2019/05/ppa-discussion-thomas-rowe-and-alex-voorhoeves-egalitarianism-under-severe-uncertainty-with-critical-precis-by-brian-jabarian/}{here}.
}}

\date{This version: \today \\
First version: May 15, 2019}

\begin{document}
\maketitle

\smallskip
\begin{abstract}
\singlespacing
\noindent In their article, ``Egalitarianism under Severe Uncertainty'', (\textit{Philosophy and Public Affairs}, 46:3, 2018), Thomas Rowe and Alex Voorhoeve develop an original moral decision theory for cases under uncertainty, called ``pluralist egalitarianism under uncertainty". In this paper, I firstly sketch their views and arguments. I then elaborate on their moral decision theory by discussing how it applies to choice scenarios in health ethics. Finally, I suggest a new two-stage Ellsberg thought experiment challenging the core of principle of their theory. In such an experiment pluralist egalitarianism seems to suggest the wrong, morally and rationally speaking, course of action -- no matter whether I consider my thought experiment in a simultaneous or a sequential setting. 
\bigskip

\end{abstract}
\smallskip
\newpage


\section{Introduction}
In \cite{rowe_egalitarianism_2018}, Thomas Rowe and Alex Voorhoeve develop an original moral decision theory for cases under uncertainty, called ``pluralist egalitarianism under uncertainty". 

In this paper, I firstly sketch their views and arguments. I then elaborate on their moral decision theory by discussing how it applies to choice scenarios in health ethics. Finally, I suggest a new two-stage Ellsberg thought experiment challenging the core of principle of their theory. In such an experiment pluralist egalitarianism seems to suggest the wrong, morally and rationally speaking, course of action -- no matter whether I consider my thought experiment in a simultaneous or a sequential setting.

Let me first introduce Rowe and Voorhoeve's theory. While one can find the original defense of pluralist egalitarianism under risk elsewhere\footnote{See \cite{fleurbaey_welfare_2018}, \cite{fleurbaey_assessing_2010}, Fleurbaey and Voorhoeve (2018) and \cite{otsuka_equality_2018}.}, in this paper, the authors focus on its uncertainty version, which they call it ``pluralist egalitarianism under uncertainty''. For sake of brevity and to differentiate the risk and uncertainty version, let me label pluralist egalitarianism under risk
{\bfseries\scshape per} and pluralist egalitarianism under uncertainty, {\bfseries\scshape peu}. Briefly, {\bfseries\scshape peu} holds that
\begin{quotation}
\singlespacing
[One] should improve people's prospects for well-being, raise total well-being, and reduce inequality in both people's prospects and in their final well-being (how well their lives end up going) (Rowe \& Voorhoeve, 2018, 243-244).
\end{quotation}
\smallskip

This paper, which focuses on this latter version, is organized as follows. In Section 2, we provide some background on the pluralist egalitarianism under uncertainty. In Section 3, we elaborate on this theory by discussing its concrete moral recommendations in the context of health ethics. In Section 4, we propose a thought experiment within their context of health ethics, challenging both the moral permissibility and rational permissibility of {\bfseries\scshape peu}.

 \section{The Permissibility of Pluralist Egalitarianism} 
 In this section, I resume the moral permissibility and the rational permissibility of {\bfseries\scshape peu}. In both cases, the authors justify the permissibility by showing that the core principle of {\bfseries\scshape peu} that I label {\bfseries\scshape the uncertainty moral aversion principle}\footnote{I use this label to avoid the confusion with the standard rational principle of ``uncertainty aversion principle'' that we present here after.} complies with specific moral principles and rational principles. Let me first start by presenting these targeted principles.
 
The targeted rational principle is derived from the orthodox rational principles for decision-making under uncertainty. These latter principles are themselves derived from orthodox rational principles for decision-making under risk. One important principle under risk, relevant to present {\bfseries\scshape peu}, is the following: under risk, the decision-maker is assumed to be capable of making up their mind to form (or has access to) precise subjective probabilities regarding the possible states of nature (\cite{savage_foundations_1972}). 

By contrast, under uncertainty, the decision-maker loses this \textit{precision} ability. Depending on the degree of severity of uncertainty, an agent will be capable to come up with more or less \textit{imprecise} probabilities. At the lowest degree of severity, the decision-maker is assumed to be able to come up with a reasonable range of probabilities (cases of ``moderate uncertainty'' in the authors' formulation). At the upper level, it becomes more difficult to form probabilities. In such a case, the agent ends up with extreme range of probabilities (cases of ``severe uncertainty''). At the highest level of severity, it becomes simply implausible to imagine an agent capable of reasoning with probabilities: the agent deals with no probability at all (cases of ``maximal uncertainty"\footnote{Also known as ``ignorance'' in the philosophical literature, see \cite{bradley_decision_2017}. It is worth noticing that the usage of these expressions varies depending on the niche literature that one reader ends up reading, either or economics and also within different philosophy niches, which might confuse the general audience. Hence, ``severe uncertainty'' could refer to ``ambiguity'', as ``moderate uncertainty'' can as well. But these misuses go beyond this specific ambiguity debate. For this reason, in this pr\'ecis, we strictly follow the authors' terminology to avoid more confusion}). For each of these degrees of severity, decision theorists from economics and philosophy have developed specific classes of decision-making models, each one differentiated by a variation of a generic standard rational principle under uncertainty: the uncertainty aversion principle.   

This is the principle that Rowe and Voorhoeve use to build {\bfseries\scshape the uncertainty moral aversion principle}. Given the fact that the authors ambition to develop a ``moral decision theory under uncertainty'', namely a decision theory which provides not only rational but also moral recommendations under uncertainty, such a the standard uncertainty aversion principle has to be adjusted such that it complies with the standard moral principles. Hence, the resulting morally adjusted uncertainty aversion principle is the {\bfseries\scshape the uncertainty moral aversion principle}. 

Before introducing this principle, let us present the standard moral principles that {\bfseries \scshape peu} principle has to comply with. The moral principle the authors look for is an egalitarian principle, different from the standard ones in circulation in the moral philosophy literature. Standard definitions of equality rely on the outcome's values: two situations are equal if and only if individuals end up equally well off. One could extend this claim under uncertainty such that the only morally relevant information to define "equality" is still the outcome's values. However, according to the authors, this definition of equality, pertaining exclusively to the value of final well-being, would rule out crucial moral information to design a fair system of distributive justice. 
 Accordingly, one should incorporate the experience of uncertainty itself in the definition of equality under uncertainty. This integration operates as a moral benefit or a cost in the system of distributive justice. In Particular, for Rowe and Voorhoeve, facing uncertainty is a "burden"  (op. cit. p. 242) in the sense of depressing the value of an individual's prospects. Therefore, it should correspond to a moral cost. Let us see why in the following situation.  Suppose Ann will go wholly blind unless she receives a treatment. As her doctor, you have two alternative treatments. The first treatment is well-known to all and risky. It has a 50\% chance of curing her and 50\% chance of not affecting her. Since you have implemented it in the past, and so, have access to a small distribution of success and failure, you have certain prior beliefs in these current objective estimations of success and failure. The second treatment is entirely new and maximally uncertain. It leads to a full cure or no cure at all, with no objective estimation of failure and success accessible. Since it is very new, you are not familiar enough to counter the absence of objective estimates by forming precise prior beliefs about its effectiveness. Despite leading to the same two possible levels of final well-being as the risky treatment, we can say that, in prospect, the uncertain treatment bears a moral cost. 

Based on these rational and moral considerations, the core principle of {\bfseries \scshape peu} is {\bfseries\scshape the uncertainty moral aversion principle} complies with both moral principles and rational principles. This latter holds that when choosing between a risky prospect and an uncertain prospect, one should opts for the risky prospect\footnote{As initially shown in Ellsberg's original thought experiment, see \cite{ellsberg_risk_1961}}. With respect to the descriptive rational permissibility, we can attest that the uncertain aversion principle is descriptively accurate since the empirical results show that this principle describes adequately how subjects behave uncertainty. However, the philosophical and economic theory literature still contests the normative rational permissibility of this uncertainty aversion attitude. Namely, the debate has not yet reached a consensus on whether a rational agent may permissibly display an uncertainty averse attitude. Rowe and Voorhoeve do not engage in this controversy. They instead adopt an assumption defended by some leading decision theorists according to which it is rationally permissible to display an uncertainty averse attitude, despite not being rationally required. This debate has focused in length on the normative rational permissibility of this attitude. 

However, much less has been said about its moral permissibility. Rowe and Voorhoeve's paper is important because it fills this gap. And, it is original in developing a specific egalitarian interpretation of the uncertainty aversion. Finally, if we take again Ann's example introduced above, one would say that granting uncertainty aversion, it would be morally impermissible to choose to incur on Ann's behalf, for anyone concerned by her welfare, which seems a sensible concern for any distributive justice theory. In sum, pluralistic, uncertainty-averse egalitarianism favors alternatives for which more fine-grained probabilistic information related to the states of nature is available.  Besides, this view considers uncertainty as important moral information to rely on to take a fair distributive justice based decision and should count as a moral cost (in the sense of depressing the value of individuals' prospects) in the system of distributive justice.

\section{Applying Pluralist Egalitarianism to Health Ethics}

Let us now see how {\bfseries \scshape peu} works in a more concrete situation. Suppose Ann and Bea will go wholly blind and with lifetime well-being of 50 (moderately good quality of life) unless they receive treatment. If fully cured, each individual would have lifetime well-being of 80 (very good quality of life). Both are strangers to the decision-maker and each other. Unfortunately, the resources at the decision-maker's disposal do not suffice to cure both Ann and Bea fully. As listed below, three classes of treatments are available to the decision-maker: certain, risky, and uncertain treatments. The classes of certain and risky treatments each contain two different treatments. The class of uncertain treatments includes four different treatments. Among these eight treatments, the decision-maker has to choose one. Hence, a distributive theory of justice can help to make up the decision-maker's mind. 

\paragraph{Options}

\begin{enumerate}[label=(\arabic*)]
    \item {\bfseries\scshape \small inequality under certainty}. Cure Ann and leave Bea to go wholly blind. 
    \item {\bfseries\scshape \small equal risk, unequal final well-being}. This treatment either cures Ann and is entirely ineffective for Bea (leaving her to go wholly blind) or, instead, is entirely ineffective for Ann (leaving her to go wholly blind) and cures Bea. These results are equally likely.
    \item {\bfseries\scshape \small equality under risk}. This treatment either cures both individuals or is wholly ineffective for both, with each result being equally likely.
    \item {\bfseries\scshape \small equality under certainty}. This treatment improves both Ann's and Bea's condition to that of a merely partial, but still substantial, visual impairment. We will consider both cases in which the level of well-being associated with this partial impairment is precisely halfway between the well-being associated with complete blindness and a full cure and cases in which this level falls short of this halfway point. The shortfall is given by a cost c, with 0 $<$ c $<$ 15.
    \item {\bfseries\scshape \small equal uncertainty, unequal final well-being}. This treatment will either cure Ann and leave Bea wholly blind or cure Bea and leave Ann wholly blind, with no information available about the probability of either outcome.
    \item {\bfseries\scshape \small equality under uncertainty}. This treatment either cures both individuals or leaves them both to go wholly blind, with no information available about the probability of either outcome.
    \item {\bfseries\scshape \small unequal uncertainty}. Ann is given a novel treatment that either cures her or instead leaves her wholly blind, with no information about the probability of either outcome. Bea is given a distinct treatment which will either, with probability 0.5, cure her or, with probability 0.5, leave her wholly blind.
    \item {\bfseries\scshape \small equal moderate uncertainty}. Ann and Bea are each given different distinct, moderately uncertain treatments, each of which will either offer a full cure or leave its recipient wholly blind. For each of their treatments, the probability of a cure ranges from 0.25 to 0.75.
\end{enumerate}

\paragraph{Pairwise Comparison of Options}

Then, the authors consider the following pairwise comparison of alternatives. They deduce the preference relations between the alternatives from their pluralist egalitarianism. We sum up briefly these resulting preference relations. 

\begin{enumerate}[label=\Alph*.]
\setlength\itemsep{-0.1em}

\item (1) {\bfseries\scshape \small inequality under certainty} \textit{versus} (2) {\bfseries\scshape \small equal risk, unequal Final well-being}. (2) is morally preferred to (1) because it gives an equal shot of being cured for Ann and Bea whereas (1) cures one of the two arbitrarily.
  
\item (3) {\bfseries\scshape \small equality under risk} \textit{versus} (1) {\bfseries\scshape \small inequality under certainty}. (3) is morally preferred to (1) because not only does it give an equal shot of being cured for Ann and Bea but it also eliminates all interpersonal unfairness but still maximizes total expected utility.

\item (3) {\bfseries\scshape \small equality under risk} \textit{versus} (2) {\bfseries\scshape \small equal risk, unequal final well-being}. (3) is morally preferred to (2) for the same reasons in B.

\item (4) {\bfseries\scshape \small equality under certainty} \textit{versus} (4) {\bfseries\scshape \small  inequality under certainty}. Suppose there is no cost to remove the inequality, say, for $c=0$. In this case, (4) is morally preferred to (1)  because inequality is suppressed without loss in expected total well-being.
    
\item (4) {\bfseries\scshape \small equality under certainty} \textit{versus} (2) {\bfseries\scshape \small equal risk, unequal final well-being}. Suppose $c=0$. (4) is morally preferred to (2) for the same reasons as in D.
    
\item (4) {\bfseries\scshape \small equality under certainty} \textit{versus}  (3) {\bfseries\scshape \small equality under risk}.
Suppose $c=0$. (4) is morally indifferent to (3) because both offers Ann and Bea equal expected well-being and both are equally good prospects for each person  and neither contains any inequality.
   
\item (4) {\bfseries\scshape \small equality under certainty} \textit{versus} (1) {\bfseries\scshape \small inequality under certainty}. For some, sufficiently small $c>0$, (4) is morally preferred to (3) because it eliminates inequality.
    
\item (4) {\bfseries\scshape \small equality under certainty} ($c>$0 but $c$ very small) \textit{versus} (3) {\bfseries\scshape \small equality under risk}. For all $c>0$ (3) is morally preferred to (4) because neither contains any inequality but (3) contains more valuable prospects for each person.
    
\item (5) {\bfseries\scshape \small equal uncertainty, Unequal final well-being} \textit{versus} (2) {\bfseries\scshape \small equal risk, unequal final well-being}. (2) is morally preferred to (5) because this latter, due to the presence of uncertainty, reduces the value of individuals' prospects.
    
\item (6) {\bfseries\scshape \small equality under uncertainty} \textit{versus} (3) {\bfseries\scshape \small equality under risk}. (3) is morally preferred to (6) because (6) reduces the value of each individual's prospects and population level value prospect. (6) generates population's level uncertainty, because the decision-maker is uncertain about the anonymized distribution of final well-being\footnote{Hence, (5) does not generate population's level uncertainty but comes at the expense of certain inequality in the final value of well-being between Ann and Bea.}.
    
\item (7) {\bfseries\scshape \small unequal uncertainty} \textit{versus} (8) {\bfseries\scshape \small equal moderate uncertainty}. (8) is morally preferred to (7) because it distributes equally ex ante an equal quantity of uncertainty over Ann and Bea and thus its moral cost is shared equally among them.
    
\item (5) {\bfseries\scshape \small equal uncertainty, unequal final well-being} \textit{versus} (1) {\bfseries\scshape \small equality under certainty}. For any decision-maker willing to incur a cost ($c>0$) to eliminate the uncertainty and/or the inequality, (1) is morally preferred to (5).
\end{enumerate}

\paragraph{Optimal {\bfseries \scshape peu} Strategy}

Despite leading to the same outcome in terms of well-being (either cured or not), the risky and uncertain treatments do not bear the same moral values. The uncertain treatment is morally more costly than the risky treatment because it would expose Ann and Bea to experiencing unnecessary uncertainty, which is morally impermissible. Overall, uncertainty aversion and inequality aversion incur a cost to remove inequality. Hence, if the decision-maker is inequality averse, but uncertainty neutral then for her, (2) and (3) are equivalent.

\section{Objections: A Two-Stage Ellsberg Thought Experiment}\label{Objections}

In this section, I propose two new thoughts experiments that I label ``two-stage Ellsberg thought experiments'' and consider them in the case of health ethics. But before presenting them, I rather show why such theoretical considerations will be relevant for applied health ethics.

Consider the following scenario. You are the medical decision-maker for Ann and you have to administer a treatment to her. This treatment is formed by the prescription of two medicines at your disposal: \textbf{\scshape medicine a} and \textbf{\scshape medicine b}. For the treatment to work, you have to prescribe two medicines to Ann but you are not obligated to prescribe only two different medicines. You may decide to prescribe both times the same medicine - suppose you have enough supply. Before giving more information regarding the efficacy of each medicine, I shall precise that I consider hereafter two versions of this decision problem. 

I firstly consider the case where you have to decide \textit{simultaneously} both medicines that Ann will take. That is to say, once she has taken the first medicine, although you observe her resulting intermediary health condition, you cannot intervene to revise your decision made regarding the second medicine that she is about taking - no matter what. Accordingly, at starting time \textit{t}, you have \textbf{four} options: 
\begin{enumerate}[label=(\Roman*)]
\item \textbf{\scshape medicine a} and then \textbf{\scshape medicine b}
\item \textbf{\scshape medicine a} and then \textbf{\scshape medicine a}
\item \textbf{\scshape medicine b} and then \textbf{\scshape medicine b}
\item \textbf{\scshape medicine b} and then \textbf{\scshape medicine a}
\end{enumerate}
At time \textit{t+1} you have no more choice to make. At time \textit{t+2}, you observe her resulting final health condition and know whether she is fully cured or not. 

Then, I consider a dynamic case. You have now to decide \textit{sequentially} both medicines. That is to say, once you have decided her first medicine and she has taken it, you observe her resulting intermediary health condition and you intervene to decide her second medicine.
Hence, at starting time \textit{t}, you have \textbf{two} options: 
\begin{enumerate}[label=(\Roman*)]
\item \textbf{\scshape medicine a}
\item \textbf{\scshape medicine b}
\end{enumerate}

Then, at time \textit{t+1} you have a new choice to make from these two \textbf{two} options:  \begin{enumerate}[label=(\Roman*)]
\item \textbf{\scshape medicine a}
\item \textbf{\scshape medicine b}
\end{enumerate}

In time \textit{t+2}, you observe her final health condition and know whether she is fully cured or not.

We now turn to the description of the information regarding the efficacy of each medicine. On the one hand, suppose that \textbf{\scshape medicine a} is well-known to all doctors and health experts. Accordingly medical decision-makers know that \textbf{\scshape medicine a} has a 50\% of success and 50\% of failure with respect to Ann's kind of illness. We say that it has a \textit{risky prospect} of success and failure (i.e., a precise probability distribution can be assigned to it).  On the other hand, suppose that \textbf{\scshape medicine b} has never been used before for Ann's kind of illness. Accordingly, no expert has information regarding its chances of success and failure. We say that \textbf{\scshape medicine b} has an \textit{ambiguous prospect} of success and failure (i.e., no precise probability distribution can be assigned to it). 

The choice problem in both cases simultaneous and sequential is the following: which options should the medical decision-maker choose given the fact that they want to maximize Ann's well-being (i.e., to fully cure her at her maximal well-being)?

For the sake of of communicating in a simple manner my objections, I will represent in a simplified representation what is meant by ``\textbf{\scshape medicine b} (resp. \textbf{\scshape medicine a}) is successful or not'' by the drawing of a ball in a urn, as it usually done in philosophical choice theory and economic decision theory. And we say that the treatment is successful if you draw two balls of the same color in urns which contain only two colors, Black and White. I build this representation on the framework proposed by Fleurbaey (2019)\footnote{I am grateful to Marc Fleurbaey to have let me access his unpublished manuscript. The original framework within which his has been built can be found in \cite{al-najjar_ambiguity_2009}.}. 
 
On the one hand, what my thought experiments have in common with Fleurbaey's one are the following elements: there are two urns, containing blacks and white balls; one risky urn where there 50 black balls and 50 white balls and one ambiguous were there are in total 100 balls but we do not its composition (whether all Black, or all White, or some of each color, to what extent); two drawings, with replacement and winning the game if match in color. The main difference between his and mine lies in the options considered and the alternatives we consider. He considers the following three options : ``drawing from the risky urn and again from the same risky urn'', ``risky and ambiguous'' and ``ambiguous and risky''. 
 
On the other hand, the crucial difference is the following. Fleurbaey considers the following three two-stage drawing options ``risky and risky'', ``risky and ambiguous'' and ``ambiguous and risky'' where he compares these options in pairwise comparison. I propose two main differences. First, I introduce the following drawing option: ``ambiguous and ambiguous'' and second, to focus exclusively to the following comparison: compare it to the main opposite option which is ``risky and risky''. It is a fundamental since it is not intuitively straightforward to understand that ``ambiguous and ambiguous'' is actually a better option, no matter what, than the option ``risky and risky'' in terms of prospects of success\footnote{The former has its success prospects minimally bounded at $\frac{1}{2}$ whereas the later has its success prospect maximally bounded at at $\frac{1}{2}$: the best case scenario of \textbf{\scshape risky and risky} corresponds to the worst case scenario of \textbf{\scshape ambiguous and ambiguous}.}.

In the following, I briefly present all the options and the pay-offs. I will turn then to our main comparison showing how it is decisive against \textbf{\scshape peu}). We also show how FLeurbaey's experiment can be a challenge to \textbf{\scshape medicine a}) depending on whether we consider the agent to be \textit{naive} or \textit{sophisticated}.
\paragraph{Options}

\begin{enumerate}[label=(\Roman*)]
\item {\bfseries \scshape risky and risky} You pick a ball from a {\scshape risky urn}, $U_{1}$ and observe its color. It is either Red with 50\% of chance or Black with 50\% chance. You look at the color and you put it back into the urn. Then, you pick again a ball from the same {\scshape risky urn}, $U_{1}$ and observe its color. It is either Red with 50\% of chance or Black with 50\% chance.

\item {\bfseries \scshape ambiguous and ambiguous} You pick a ball from an {\scshape ambiguous urn}, $U_{2}$ and observe its color. It is either Red or Black with unknown chance. You look at the color and you put it back into the urn. Then, you pick again a ball from the same {\scshape ambiguous urn}, $U_{2}$ and observe its color. 

\item {\bfseries \scshape ambiguous and risky}	You pick a ball from an {\scshape ambiguous urn}, $U_{2}$ and observe its color. It is either Red or Black with unknown chance. You look at the color and you put it back into the urn. Then, you pick a ball from the {\scshape risky urn}, $U_{1}$ and observe its color. It is either Red with 50\% of chance or Black with 50\% chance. 

\item {\bfseries \scshape risky and ambiguous} You pick a ball from a {\scshape risky urn}, $U_{1}$ and observe its color. It is either Red with 50\% of chance or Black with 50\% chance. You look at the color and you put it back into the urn. You pick a ball from the {\scshape ambiguous urn}, $U_{2}$ and observe its color. It is either Red or Black with unknown chance. 
\end{enumerate}

\paragraph{Pay-offs}

\begin{enumerate}[label=(\Alph*)]
\item If you choose (I) and if you are successful, then Ann is cured and she ends up with a lifetime well-being of 50. She is not cured otherwise and lives a miserable life. 
\item If you choose (II) and if you are successful, then Ann is cured and she ends up with a lifetime well-being of 80. She is not cured otherwise and lives a miserable life. 
\item If you choose (III) and if you are successful, then Ann is cured and she ends up with a lifetime well-being of 60. She is not cured otherwise and lives a miserable life. 
\item If you choose (IV) and if you are successful, then Ann is cured and she ends up with a lifetime well-being of 80. She is not cured otherwise and lives a miserable life. 
\end{enumerate}

\paragraph{Optimal strategies}
In both cases, whether your two drawings are made sequential or simultaneously, the optimal strategies is (IV) {\bfseries \scshape ambiguous and ambiguous} and, contrary to the common intuition, (I) {\bfseries \scshape risky and risky} is actually the worst option. Indeed, urn Risky wins with probability $.5$. If urn Ambiguous has proportion $p$ of red balls, then it wins with probability $p^2 + (1-p)^2 \geq .5$
and it is strictly superior when $p \neq .5$.

\paragraph{What {\bfseries \scshape peu} recommends}
According to {\bfseries \scshape peu} pluralistic, uncertainty-averse egalitarianism, you should be uncertainty averse and therefore reject (IV) - as would advise you, I suspect, any model of decision-making under ambiguity, in a weakly preferred sense.

\paragraph{Being ambiguity averse in sequential drawing is also irrational}

First, let see the irrational way to uncertainty averse, which pluralistic uncertainty-averse egalitarianism could advice you, but clearly you should not follow in a sequential setting. This way corresponds to being averse to uncertainty and a non-expected utility maximizer. In this context, you can be either a naive non-expected utility maximizer or a sophisticated non-expected utility maximizer.

On the one hand, let us the consider the case where the decision-maker is naive. By naive, I mean that the decision-maker compares only options available during the first period, \textit{t} and does not anticipate the comparison of the future options at time \textit{t+1}.

At time \textit{t} you compare two options: drawing from {\scshape ambiguous urn} or drawing from. Since, following {\bfseries \scshape peu}, the experience of of severe uncertainty should count as a moral cost, you discard drawing from {\scshape ambiguous urn} and prefer drawing from {\scshape risky urn}. You look at the color and put the ball back. 

Once the first drawing is done, at time \textit{t+1}, you learn that you have two options that you had not foreseen, drawing from again {\scshape risky urn} or {\scshape ambiguous urn}. Again, according to {\bfseries \scshape peu} you should decide to draw from {\scshape risky urn}. 

In resume you went with {\bfseries \scshape risky and risky} but it is dominated by {\bfseries \scshape ambiguous and risky}. If you had thought twice, you would have been better off by going directly at first with {\bfseries \scshape ambiguous and risky}. And if you had thought one more, you would have been maximally better off by going with {\bfseries \scshape ambiguous and ambiguous}. But you could not go neither with {\bfseries \scshape ambiguous and risky} nor with {\bfseries \scshape ambiguous and risky} because you followed the uncertainty aversion principle and had limited rationality. This irrationality can be viewed as a dynamic inconsistency. 

On the other hand, let me consider the case where you are a sophisticated decision-maker. 

At \textit{t}, you anticipate the choice you will face at time \textit{t+1}. You depress the value of option {\bfseries \scshape risky and ambiguous} because of the presence of ambiguity at the second stage. And since for the option {\bfseries \scshape risky and risky}, the second drawing is dominated by {\bfseries \scshape ambiguous and risky}, you decide to draw from {\bfseries \scshape ambiguous and risky}, where finally you have 50\% chances to have two colors of the same urn (since the ambiguous urn is followed automatically by a risky urn, the ambiguity in the first round has no negative effect on you). You cure Ann but not fully. Despite not being dynamic inconsistent, the sophisticated agent violates the Independence of Irrelevant Alternatives.

For these reasons, the moral permissibility of a distributive theory of justice designed for society as a whole (and not for each individual), based on the violations of such rationality principles, should be rejected. This could be the original recommendation emerging from Rowe and Voorhoeve's use of the Hurwicz criterion, independently from expected utility theory. 

\paragraph{Being ambiguity averse in sequential is unreasonable}

Second, let see the unreasonable way to be uncertainty averse, which pluralistic uncertainty-averse egalitarianism could also advice you. This way is compatible with being an expected-utility maximizer. In this context, you will choose option {\bfseries \scshape ambiguous and risky} over option {\bfseries \scshape risky and risky} because the latter is dominated by option {\bfseries \scshape ambiguous and risky}. Besides, the cost of experiencing uncertainty in {\bfseries \scshape risky and ambiguous} is so high that its expected value would be lower than the expected value of option {\bfseries \scshape ambiguous and risky}. Therefore you go with option {\bfseries \scshape ambiguous and risky}. This is the option that pluralistic egalitarianism that recommends, by embedding the Hurwicz criterion with expected utility\footnote{One might look at \cite{gul_hurwicz_2015},\cite{binmore_rational_2008} and \cite{binmore_minimal_2016} for such kind of a model: the main difference with the original Hurwicz's criterion \cite{hurwicz_optimality_1951} is that here the concept of uncertainty is decomposed as \textit{perceived uncertainty} and \textit{source uncertainty attitude}. The latter corresponds to the original Hurwicz's criterion that Rowe and Voorhoeve integrate into their definition of equality under uncertainty, but the former seems not mobilized and integrated in the value functions of their pluralistic egalitarianism.}. This alternative might be unreasonable compared to option {\bfseries \scshape risky and ambiguous}.

If the decision-maker has access to Ann's information with regards to her attitude, and she is willing to go with {\bfseries \scshape risky and ambiguous}, then there is no reasonable justification, as for me, to overcome her attitude by imposing a cautious attitude from the social planner on Ann and thus depriving her from an extra lifetime well-being (gained from 9a). From an aggregation perspective, the correct uncertainty attitude that the social planner should take into account is the one emerging from the citizens, except if she has another better - qualitatively speaking - source of information. But in any event, imposing one's own attitude for the entire society seems, at least, questionable and the same for all public policy decisions, even more (evidence might show that citizens are uncertainty averse towards specific public decisions and uncertainty seeking towards others).

\section{Concluding Remarks}

From this paper, one take away is that the Hurwicz criterion, embedded in the expected utility approach (let's note it ``HEU'') seems to be flexible enough to accommodate different levels of uncertainty attitudes depending on the context of risk and severe uncertainty. For instance, if maximally uncertainty averse HEU agents ($\alpha = 1$) always prefer bets on less uncertain sources, HEU agents with an intermediate level of uncertainty aversion ($0 < \alpha < 1$) reverse this preference when the probability of winning is low see (\cite{gul_hurwicz_2015}.

Finally, I would like to point out some directions we could further discuss in the comments. Before drawing out one objection below, I rather prefer to suggest four questions to discuss.  

First, should decision-makers always be uncertain aversion (at the societal level) when the expected well-being prospects under ambiguity of individuals (at the individual-level) are very high? 

Second, for any distributive theory of justice under severe uncertainty (or ambiguity), should we consider in priority the experience of uncertainty as an impermissible moral cost and avoid it at all expenses, even if this avoidance lead to the violation of rational principles or not? 

Third, when social planners \textit{do not have} information regarding  citizens' attitudes towards uncertainty, which option should they take between a severely uncertain option with high expected pay-offs and moderately uncertain option with low expected pay-offs for individuals? Suppose they opt for the less uncertain option, following the recommendations of cautious expected utility {\bfseries \scshape ceu}  (\cite{cerreia-vioglio_cautious_2015}, in line with those of {\bfseries \scshape peu}. Should they always use this this theory or is there any case where it would be better for individuals' welfare not to follow these latter theories? 

Fourth, when social planners \textit{have} information regarding citizens' attitudes towards uncertainty, do they have the moral authority and political legitimacy to override this information if citizens' attitudes contradict the recommendations of {\bfseries \scshape ceu} and {\bfseries \scshape peu} and always follow the recommendation of these latter theories or not ?

\bibliography{EllsbergPhilo.bib}

\end{document}